\documentclass[a4paper,12pt]{article}
\usepackage{amsmath}
\usepackage{amssymb}
\usepackage{listings}
\usepackage{comment}
\usepackage[pdftex]{graphicx}
\newcommand {\be} [1]   {\begin{equation}\label{#1}}
\newcommand {\ee}       {\end{equation}}

\def\({\left(}
\def\){\right)}
\newcommand {\eq} [1]   {(\ref{#1})}
\newcommand {\DS} {\displaystyle}
\begin{document}
\bigskip
\begin{center}
{\LARGE
Structural model for the dynamic buckling of a column under constant rate compression}
 \\ [4mm]
\large
Vitaly A. Kuzkin\footnote{kuzkinva@gmail.com}
 \\ [4mm]
 Institute for Problems in Mechanical Engineering RAS \\ [4mm]
  Saint Petersburg Polytechnical University \\ [4mm]
\end{center}
\bigskip
\bigskip
\begin{abstract}
Dynamic buckling behavior of a column~(rod, beam) under constant rate compression  is considered. The buckling is caused by prescribed motion of column ends toward each other with constant velocity. Simple model with one degree of freedom simulating static and dynamic buckling of a column is derived. In the case of small initial disturbances the model yields simple analytical dependencies between the main parameters of the problem: critical force, compression rate, and initial disturbance. It is shown that the time required for buckling is inversely proportional to cubic root of compression velocity and logarithmically depends on the initial disturbance.
Analytical expression for critical buckling force as a function of compression velocity is derived. It is shown that in a range of compression rates typical for laboratory experiments the dependence is accurately approximated by a power law with exponent close to~$2/3$. Theoretical findings are supported by available results of laboratory experiments.

\end{abstract}

\section{Introduction}
Buckling of columns~(rods, beams) under compression is a classical problem for mechanics of solids. In 1744 Leonard Euler predicted critical buckling force for compressed column in statics. Numerous experimental and theoretical studies have revealed that behavior of a column in dynamics is significantly more complicated. In particular, in dynamics the maximum force acting in the column, further refereed to as critical force, exceeds Euler force. Dynamic buckling behavior significantly depends on the way of compression. A review of different loading conditions may be found, for example, in a review paper~\cite{Karagiozova}. Sudden application of a force was investigated, for example, in recent works~\cite{Belyaev, Morozov_Tovstik}. Mass falling on the rod was studied theoretically and experimentally in papers~\cite{Waas, Mimura 2008}.  In laboratory experiments the load-bearing capacity of columns is usually measured in hydraulic testing machines. In this case column ends move toward each other with prescribed velocity~\cite{Hoff, Mimura 2012}. This loading regime is also typical for computer simulation of buckling at macro~\cite{Motamarri} as well as micro and nano scale levels~\cite{nanutbes 2012}. In 1951 Hoff has proposed the following simplified statement of this problem~\cite{Hoff}. Compression of a column with initial imperfection in a form of a sine wave was considered. It was assumed that the deflection of a column has the shape of the first buckling mode. This assumption yields the nonlinear differential equation describing buckling of a column. In particular, Hoff has shown that the critical force strongly depends on compression rate and initial imperfection of a column. Buckling under constant rate compression in Hoff's statement was studied in papers~\cite{Motamarri, Sevin 1960, Dym 1968, Elishakoff, Kounadis}. The influence of axial inertia~\cite{Sevin 1960}, random imperfection~\cite{Elishakoff} and boundary conditions~\cite{Motamarri} was investigated. Approximate solutions of Hoff's equation are discussed in papers~\cite{Dym 1968, Kounadis}. The dependence of critical buckling force on the rate of compression is obtained by fitting the numerical numerical results~\cite{Motamarri}. Recent experiments~\cite{Mimura 2012} has revealed that the dependence is accurately approximated by the power law. However to our knowledge theoretical explanation of this fact is not present in the literature.

In the present paper simple one-dimensional model of dynamic buckling under constant rate compression is presented. It is assumed that the equilibrium configuration of the column is perfectly straight. Understanding of buckling of perfect columns is especially important at nano scale~\cite{nanutbes 2012, nanowires power law}, because nano-objects may have no defects. The disturbance is introduced by non-zero initial deflection. At nano scale it might be associated with thermal motion. Analytical dependencies of the time required for buckling and critical force on compression rate are derived. Theoretical findings are supported by the results of laboratory experiments~\cite{Mimura 2012}.

\section{One-dimensional model, analytical solution}
\begin{figure*}[htb]%
\begin{center}
\includegraphics*[scale=0.15]{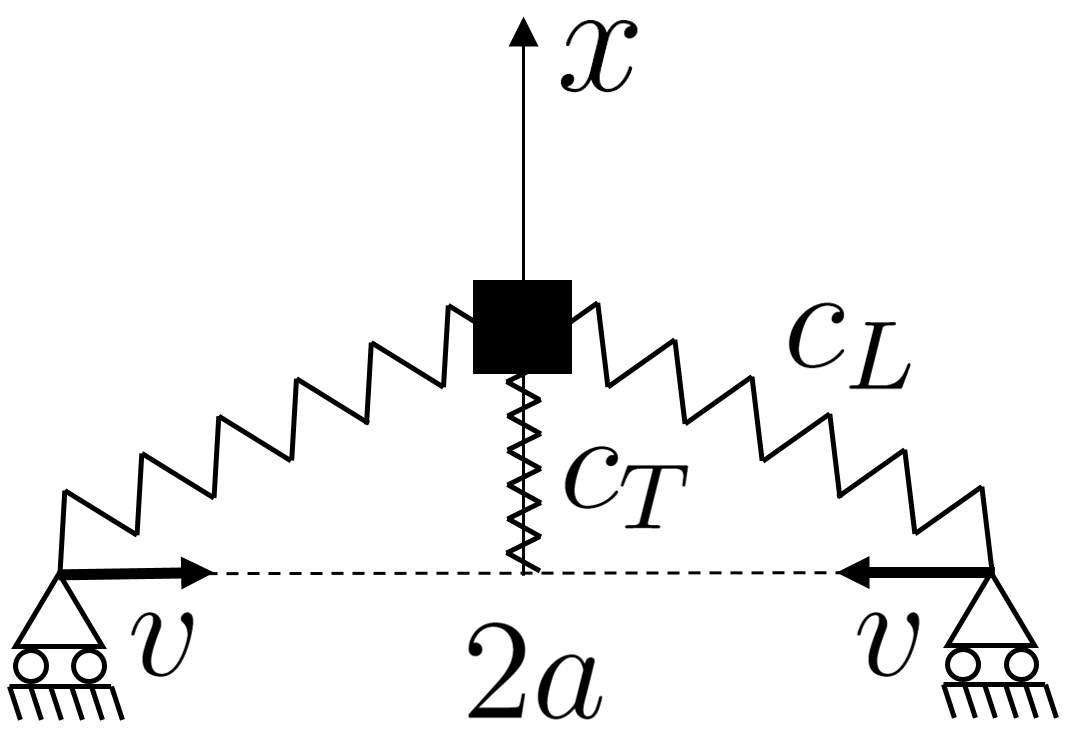}
\caption{Structural model for dynamic buckling of a column  under constant rate compression.}
\label{fig1}
\end{center}
\end{figure*}

Consider simple structural model reflecting the main physical features of a column subjected to constant rate compression. The column is simulated by a particle~(point mass) connected with two walls by linear springs
with stiffness~$c_L$ and equilibrium length~$a_0$~(see figure~\ref{fig1}).
Transverse stiffness of the column is
simulated by a spring with stiffness~$c_{T}$ connecting the particle with middle point between the walls~(see figure~\ref{fig1}).
The walls move toward each other with constant velocity~$v$.
Potential energy of the system has the form:
\be{}
  U = c_L(\sqrt{a^2 + x^2} - a_0)^2 + \frac{1}{2}c_T x^2,
\ee
where~$x$ is a coordinate of the particle; $a, a_0$ are the current and initial half distances between the walls.

For any distance between the walls~$2a$ the straight configuration of the system~$x=0$ is an equilibrium.
At some distance between the walls this equilibrium  becomes unstable
and two additional stable equilibria exist:
\be{new_eq}
 x_{1,2} = \pm \sqrt{a_E^2 - a^2}, \quad a \leq a_E, \qquad  a_E = \frac{2c_l a_0}{2c_L + c_{T}}.
\ee
Here~$a_E$ is a bifurcation point corresponding to critical half-distance between the walls. It can be show that for~$a>a_E$ the straight configuration on the system~($x=0$) is unstable.
Stable equilibria are defined by formula~\eq{new_eq}. Corresponding Euler-like critical force in statics is
\be{Euler-like}
  F_E = c_L a_0 \varepsilon_E, \qquad   \varepsilon_E = 1 -\frac{a_E}{a_0} =  \frac{c_T}{2c_L + c_T},
\ee
where~$\varepsilon_E$ is a critical deformation in statics. 
Thus static behavior of the system is qualitatively similar to the behavior of the column under compression. Below
 critical deformation the straight configuration of the system is stable. For higher deformations there are two symmetric equilibria~\eq{new_eq}. Critical value of the force corresponding to static buckling is given by formula~\eq{Euler-like}.

Consider the influence of dynamic effects on critical buckling force. Only transverse motions of the particle are considered. The effect of axial inertia is small
and can be neglected~\cite{Sevin 1960}.  The equation of transverse motion of the particle is
\be{EM}
 m\ddot{x} = -\(2c_L + c_T -  \frac{ 2c_L a_0}{\sqrt{a^2 + x^2}}\) x, \quad x(-t_E)=x_0, \quad \dot{x}(-t_E) = 0, \qquad a = a_E - vt.
\ee
where~$m$ is mass of the particle,  $x_0$ is an initial deflection. Loading is applied at~$t=-t_E$, where~$t_E=(a_0-a_E)/v$ is a time required for reaching Euler static force. Initial disturbance is simulated by non-zero initial deflection~$x_0$. Note that the statement used by Hoff is different~\cite{Hoff}. In paper~\cite{Hoff} it is assumed that natural configuration of the column has finite curvature. In the present paper perfectly straight column is considered.

Horizontal component of the force~(see fig.~\ref{fig1})
acting on the walls has the form:
\be{Fmax}
  F = c_L (a_E-vt)\(\frac{a_0}{\sqrt{(a_E-vt)^2+x^2}}-1\).
\ee
The force~$F$ has maximum value at moment of time~$t_{*}$, further refereed to as time required for buckling~\cite{Waas}. Parameter~$t_*$ is defined by the equation:
\be{tstar}
 \frac{{\rm d}}{{\rm d} t} F(x(t, v, x_0), vt)|_{t=t_*} = 0\quad\Rightarrow\quad t_* = t_*(v, x_0).
\ee
Equation~\eq{tstar} yields the dependence of~$t_*$ on $v$ in the implicit form. Substituting~$x(t_*)$ and~$t_*(v, x_0)$ into equation~\eq{Fmax} yields the expression for the critical buckling force:
\be{f_cr}
%F_* = c_L a_*\(\frac{a_0 }{\sqrt{a_*^2+x_*^2}}  - 1\), \qquad t_* = t_*(v, x_0), \quad x_* = x(t_*, v, x_0), \quad a_* = a_E - vt_*.
F_* = c_L (a_E - vt_*) \(\frac{a_0 }{\sqrt{(a_E - vt_*)^2+x_*^2}}  - 1\), \qquad  x_* = x(t_*, v, x_0).
\ee
The dependencies~$t_*(v, x_0)$, $x(t_*, v, x_0)$ are given in the implicit form by nonlinear equations~\eq{EM}, \eq{tstar}.

The exact solution of equations~\eq{EM}, \eq{tstar} in the general case is not
straightforward. Therefore the following assumptions are used
\be{assumptions}
   vt_* \ll a_E, \quad |x_*| \ll a_E.
\ee
Linearize the expression for critical force with respect to small parameters~$x_*$ and~$v t_*$:
\be{Fstar}
  F_* \approx F_E + c_L v t_* - c_L\frac{a_0 x_*^2}{2a_E^2}.
\ee
The relation between~$t_*$ and $v$ is derived as follows. The force~$F$ is expanded into series with
respect to~$vt$ and~$x$ and  then substituted into~\eq{tstar}:
\be{dFdt1}
\frac{{\rm d}F}{{\rm d} t}|_{t=t_*} \approx c_L\(v - \frac{a_0}{a_E^2} x_* \dot{x}_*\) = 0.
\ee
Critical values of displacement~$x(t_*)$ and velocity~$\dot{x}(t_*)$ are calculated using the equation of motion~\eq{EM}.
The latter is linearized under assumptions~\eq{assumptions}:
\be{EM_lin}
 \ddot{x} -\alpha v t x = 0, \qquad \alpha = \frac{2c_L a_0}{m a_E^2}.
\ee
The equation~\eq{EM_lin} is usually referred to as Airy's equation~(see e.g.~\cite{Airy function}). The solution of equation~\eq{EM_lin} is a linear combination of Airy's functions~${\rm Ai}$ and ${\rm Bi}$~\cite{Airy function}:
\be{sol Airy}
\begin{array}{l}
  \DS x(t) = C_1 {\rm Ai}\(t \alpha^{\frac{1}{3}} v^{\frac{1}{3}}\) + C_2 {\rm Bi}\(t \alpha^{\frac{1}{3}} v^{\frac{1}{3}}\), \qquad \tau_E = \alpha^{\frac{1}{3}}(a_E-a_0)v^{-\frac{2}{3}}, \\[4mm]
  \DS
  C_1 =  \frac{x_0{\rm Bi'}(\tau_E)}{{\rm Bi'}(\tau_E){\rm Ai}(\tau_E)-{\rm Bi}(\tau_E){\rm Ai'}(\tau_E)}, \quad
C_2=-
 \frac{x_0{ \rm Ai'}(\tau_E)}{{\rm Bi'}(\tau_E){\rm Ai}(\tau_E)-{\rm Bi}(\tau_E){\rm Ai'}(\tau_E)}.
\end{array}
\ee
For positive arguments~${\rm Ai}$ exponentially decays and~${\rm Bi}$ exponentially grows~\cite{Airy function}.
In particular, for large values of the argument the following asymptotic formulas can be used:
\be{Bi asymptote}
{\rm Ai}(z) \sim \frac{e^{-\frac{2}{3}z^{\frac{3}{2}}}}{2\sqrt{\pi} z^{\frac{1}{4}}},
\quad
 {\rm Ai}'(-z) \sim \frac{\cos\(\frac{2}{3}z^{\frac{3}{2}} + \frac{\pi}{4}\)z^{\frac{1}{4}}}{\sqrt{\pi} },
 \quad
 {\rm Bi}(z) \sim \frac{e^{\frac{2}{3}z^{\frac{3}{2}}}}{\sqrt{\pi} z^{\frac{1}{4}}}.
\ee
Therefore the first term in the expression~\eq{sol Airy} for~$x(t)$ is neglected. Then asymptotic formulas~\eq{Bi asymptote} and formula~\eq{dFdt1} yields
the following implicit relation between~$t_*$ and~$v$:
\be{tau_st}
    \(\frac{v}{v_s}\)^{\frac{2}{3}} =  \frac{\pi^2  {\rm Ai'(\tau_E)^2}}{ (1-\varepsilon_E)^{\frac{8}{3}}} \(\frac{x_0}{a_0}\)^2
        {\rm Bi}\(\tau_*\){\rm Bi}'\(\tau_*\), \qquad \tau_* = t_* \alpha^{\frac{1}{3}} v^{\frac{1}{3}},
\ee
where~$v_s = a_0\sqrt{2c_L/m}$ is a characteristic velocity, corresponding to velocity of longitudinal waves in the column.  For small initial disturbances~$x_0^2$ and finite compression rates~$v$ parameter~$\tau_*$, defined by formula~\eq{tau_st}, is large.
Then using asymptotic formulas~\eq{Bi asymptote} in~\eq{tau_st} yields the explicit dependence of the time required for buckling~$t_*$
on the compression rate:
\be{tstar_ex}
  \frac{t_*}{t_s} = g(v)\(\frac{v_s}{v}\)^{\frac{1}{3}}, \qquad g(v) = \(\frac{1-\varepsilon_E}{2} \ln\frac{v a_0^{3}(1-\varepsilon_E)^4}{\pi^{\frac{3}{2}} v_s x_0^3 {\rm Ai'}(\tau_E)^3} \)^{\frac{2}{3}}.
\ee
where~$t_s=a_0/v_s$. For small initial disturbances~$x_0^2$ and finite velocities~$v$ the dependence~$g(v)$ is weak. Therefore the time required for buckling is inversely proportional to cubic root of the compression rate.

Consider the contribution of~$vt_*$ and $x_*^2$ to the critical force~\eq{Fstar}.
For large~$\tau_*$ formula~\eq{tau_st} yields:
\be{}
 \frac{vt_* a_0}{x_*^2} \sim \frac{\tau_* {\rm Bi}'\(\tau_*\)}{{\rm Bi}\(\tau_*\)} \gg 1.
\ee
Then the last term in the formula~\eq{Fstar} can be neglected.

Substitution of formula~\eq{tstar_ex} into equation~\eq{Fstar} yields the dependence 
of the critical  force of the compression rate:
\be{powerlaw}
  \frac{F_*}{F_E} = 1 +  \frac{g(v)}{\varepsilon_E}  \(\frac{v}{v_s}\)^{\frac{2}{3}}.
\ee
Estimate critical deformation~$\varepsilon_E$ for a column. For Bernoulli-Euler beam with square cross-section, thickness~$h$ and
length~$l$ the expression for~$\varepsilon_E$ has the form:
\be{eps_E vs hl}
\varepsilon_E = \frac{\pi^2}{12} \(\frac{h}{l}\)^2.
\ee
Formulas~\eq{powerlaw} and \eq{eps_E vs hl} yield the expression for critical force:
\be{powerlaw2}
  \frac{F_*}{F_E} = 1 +  \frac{12g(v)}{\pi^2}  \(\frac{l}{h}\)^p \(\frac{v}{v_s}\)^q, \qquad p = 2, \qquad q=\frac{2}{3}.
\ee
The dependence of~$g$ on $v$ and $l/h$ is relatively weak. Therefore critical force and compression rate are related by the power law dependence. This result is in a good agreement with experimental data~\cite{Mimura 2008, Mimura 2012}. In paper~\cite{Mimura 2012} compression of columns with constant rate in a hydraulic machine was investigated. In paper~\cite{Mimura 2008} buckling under the action a heavy  body falling on the end of the column was considered. In both cases the dependence of critical force on the compression rate is approximated by formula, similar to~\eq{powerlaw}. The following values of parameters~$p,q$ were obtained: $p=2.0, q=0.68$~\cite{Mimura 2008} and $p=1.61, q=0.62$~\cite{Mimura 2012}.

Estimate the interval of compression rates, where formula~\eq{powerlaw} is applicable. The derivation is based on the assumption~$C_2 \neq 0$. It is straightforward to show that~$C_2=0$ at some points of the interval~$\tau_E < -1$. 
Then compression rate should satisfy the following condition:
\be{cr}
  \frac{v}{v_s} > \frac{\varepsilon_E^{\frac{3}{2}}}{1-\varepsilon_E}.
\ee
Additionally, the assumption~$\tau_* > 1$ was used. 
Using this condition in formula~\eq{tau_st} yields:
\be{cr2}
  \frac{v}{v_s}  >  \frac{\pi^3  {\rm Ai'(\tau_E)^3}}{ (1-\varepsilon_E)^4} \(\frac{x_0}{a_0}\)^3
        {\rm Bi}\(1\)^{\frac{3}{2}}{\rm Bi}'\(1\)^{\frac{3}{2}}.
\ee
Thus formula~\eq{powerlaw} describes the behavior of the system at velocities and initial disturbances satisfying conditions~\eq{cr} and \eq{cr2}.

\section{Comparison with numerical solution}
The model described above involves significant simplifications. In the present section the analytical results are compared with numerical solution of the exact nonlinear equation~\eq{EM}. The equation of motion~\eq{EM} is solved numerically using Verlet symplectic integration scheme. Initial velocity is equal to zero, initial displacement is equal to~$x_0$. The following values of dimensionless parameters are used~$c_T/c_L = 1.64 \cdot 10^{-6}$, $\Delta t/t_s = 10^{-2}$. The given ratio of stiffnesses corresponds to the column with length/thickness ratio about~$10^3$~(see formula~\eq{eps_E vs hl}). Velocity belongs to the interval~$v/v_s \in [10^{-8}; 10^{-3}]$ considered in laboratory experiments~\cite{Mimura 2012}.

\begin{figure*}[htb]%
\begin{center}
\includegraphics*[scale=0.4]{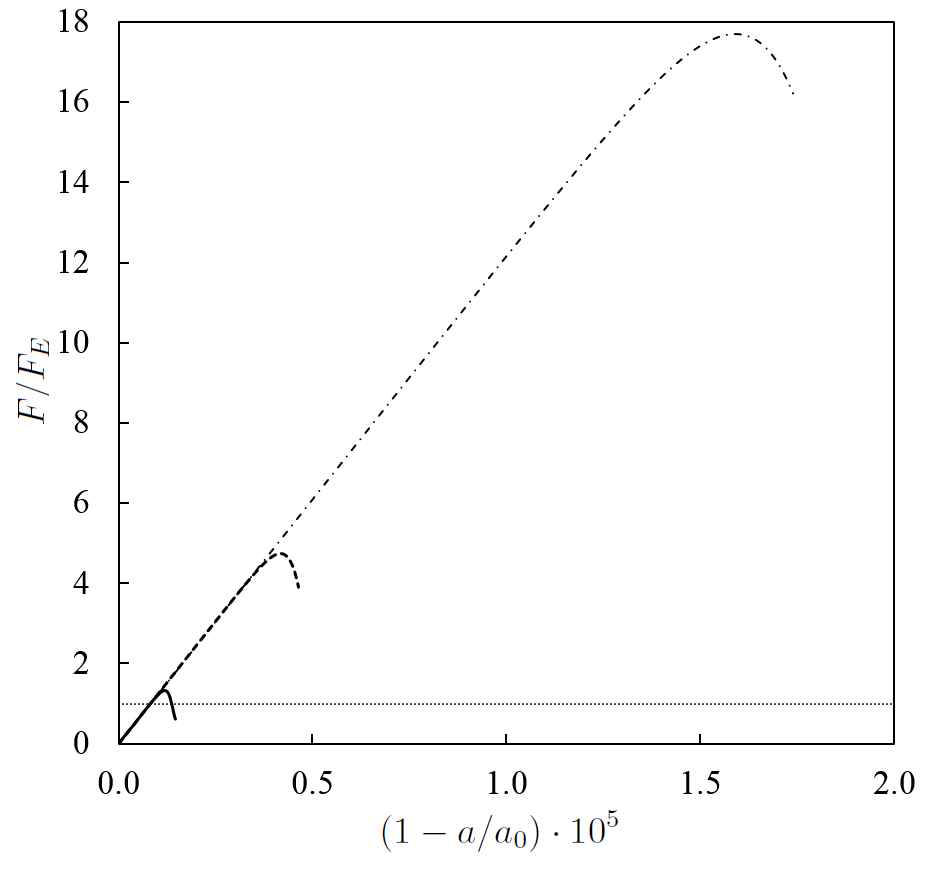}
\caption{Dependence of longitudinal force on deformation for~$x_0/a = 10^{-4}$, ~$v/v_s = 10^{-8}$~(dash-dot line), $10^{-9}$~(dashed line), $10^{-10}$~(solid line). Horizontal line corresponds to the Euler static force.}
\label{fig2_force}
\end{center}
\end{figure*}

In the simulations the longitudinal force is computed using formula~\eq{Fmax}. Typical dependence of the force on deformation for~$x_0/a = 10^{-4}$ and $v/v_s = 10^{-10}, 10^{-9}, 10^{-8}$ is shown in figure~\ref{fig2_force}.
Critical force~$F_*$ corresponds to the maximum value of the force shown in figure~\ref{fig2_force}.
Though the velocity is much lower then the ``velocity of sound''~$v_s$, the critical force significantly exceeds Euler static force~\eq{Euler-like}.

\begin{figure*}[htb]%
\begin{center}
\includegraphics*[scale=0.4]{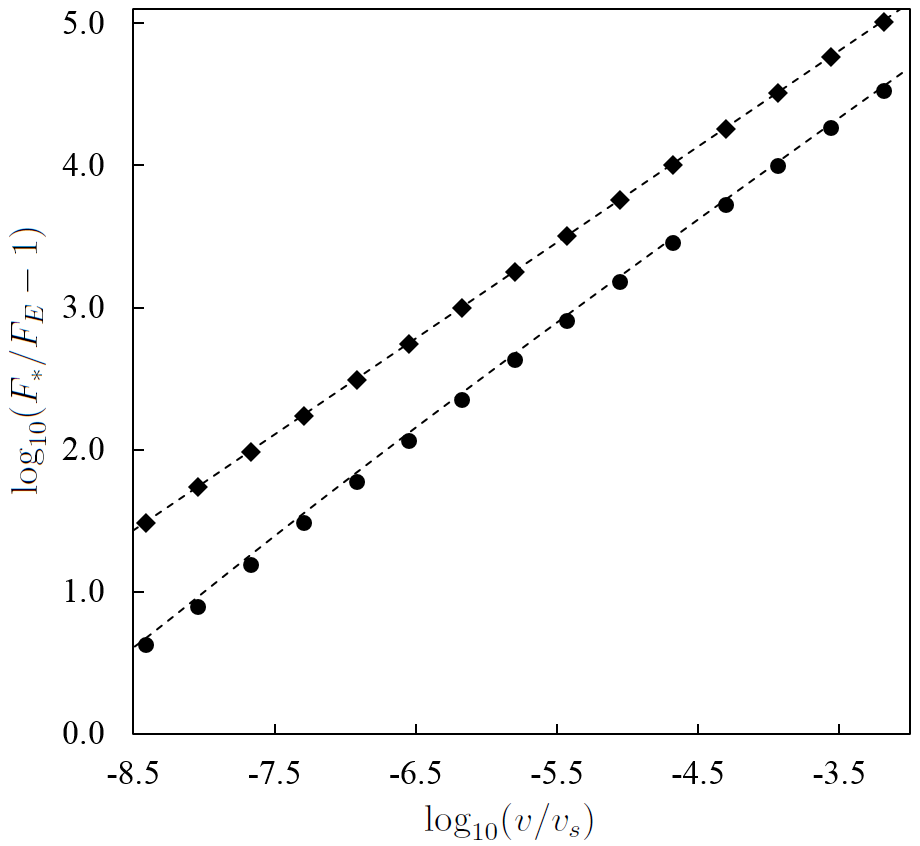}
\caption{Dependence of critical buckling force on compression rate for initial disturbances~$x_0/a_0=10^{-3}$~(circles), $10^{-12}$~(diamonds). Points correspond to numerical solution, lines correspond to analytical solution~(formula~\eq{powerlaw}).}
\label{fig3_cr_force}
\end{center}
\end{figure*}

The dependence of critical force on the rate of compression~$v$ for initial disturbances~$x_0/a_0=10^{-3}, 10^{-12}$
is shown in figure~\ref{fig3_cr_force}.
It is seen that the dependence is accurately approximated by the power law~\eq{powerlaw}. The power tends to~$2/3$ as the initial disturbance tends to zero.

\section{Conclusions}
Simple one-dimensional model for dynamic buckling of a column under constant rate compression is derived. Despite the simplicity, the model qualitatively  the influence of the main parameters~(compression rate, length/thicknessration, and initial disturbance) on the critical force. In dynamics the critical force buckling exceeds the Euler force even at relatively slow compression rates. The reason is that some time is required  for transition from unstable straight configuration to stable bend configuration. It is shown that this time is inversely proportional to cubic root of velocity of column end. For small initial disturbances the dependence of critical force on velocity of the end is accurately approximated by the  power law with exponent approximately equal to~$2/3$. This result is in a good agreement with the results of laboratory experiments~\cite{Mimura 2008, Mimura 2012}.


\begin{thebibliography}{1}
\bibitem{Karagiozova} D. Karagiozova, M. Alves, Dynamic elastic-plastic buckling of structural elements: A Review. Applied Mechanics Reviews, 2008, Vol. 61.

\bibitem{Belyaev} A.K. Belyaev, D.N. Il'in, N.F. Morozov,  Dynamic approach to the Ishlinsky–Lavrent'ev problem, Mech. Sol., 2013, Vol. 48, No. 5, pp. 504--508.

\bibitem{Morozov_Tovstik} N.F. Morozov, P.E. Tovstik, T.P. Tovstik, Again on the Ishlinskii--Lavrentyev problem, Doklady Physics, 2014, Vol. 59, No. 4, pp. 189--192.

\bibitem{Waas} W. Ji, A.M. Waas, Dynamic bifurcation buckling of an impacted column. Int. J. Eng. Sci., 2008, Vol. 46, pp. 958–-967.

\bibitem{Mimura 2008} K. Mimura, T. Umeda, M. Yu, Y. Uchida, H. Yaka, Effects of impact velocity and slenderness ratio on dynamic buckling load for long columns. Int. J. Mod. Phys. B, Vol. 22, Nos. 31, 32, 2008, pp. 5596-–5602.

\bibitem{Hoff} Hoff N.J. The dynamics of the buckling of elastic columns. J. Appl. Mech. 1951, Vol. 18, pp. 68–-74.

\bibitem{Mimura 2012} K. Mimura, T. Kikui, N. Nishide, T. Umeda, I. Riku, H. Hashimoto Buckling behavior of clamped and intermediately-suported long
rods in the static-dynamic transition velocity region. J. Soc. Mat. Sci., Vol. 61, No. 11, pp. 881---887, 2012.


\bibitem{Motamarri} P. Motamarri, S. Suryanarayan,
Unified analytical solution for dynamic elastic buckling of beams for various
boundary conditions and loading rates. Int. J. Mech. Sci., 56, 2012, pp. 60-–69.


\bibitem{nanutbes 2012} H. Shima, Buckling of Carbon Nanotubes: A State of the Art Review. Materials 2012, 5, pp. 47-84.

\bibitem{nanowires power law} C.Y. Tang, L.C. Zhang, K. Mylvaganam. Rate dependent deformation of a silicon nanowire under uniaxial compression:
Yielding, buckling and constitutive description. Comp. Mat. Sci. 51, 2012, pp. 117-–121.

\bibitem{Sevin 1960} E. Sevin On the elastic bending of columns due to dynamic axial forces including effects of axial inertia. J. Appl. Mech., 1960, 27(1), pp. 125-131.

\bibitem{Dym 1968} Dym C.L., Rasmussen M.L. On a perturbation problem in structural dynamics, Int. J. of Non-Linear Mech., 3 (2), 1968, pp. 215-225

\bibitem{Elishakoff} I. Elishakoff, Hoff's Problem in a Probabilistic Setting. J. App. Mech., 1980, Vol. 47.

\bibitem{Kounadis} A. N. Kounadis, J. Mallis, Dynamic stability of initially crooked columns under a time-dependent axial displacement of their support. Q. J. Mech. Math., Vol. 41, 4, 1988.

%\bibitem{Hutchinson_1966} J.W. Hutchinson, B. Budiansky, Dynamic buckling estimates. AIAA Journal, Vol 4, No.2, 1966

\bibitem{Airy function} F.W.J. Olver, D.W. Lozier, R.F. Boisvert, W. Clark, NIST Handbook of Mathematical Functions, Cambridge University Press, 2010.


\end{thebibliography}
\end{document}